\documentclass[%
reprint,
amsmath,
amssymb,
prb,
]{revtex4-2}
\usepackage{CJK}
\usepackage{multirow}
\usepackage{booktabs}
\usepackage{graphicx}
\usepackage{dcolumn}
\usepackage{bm}
\usepackage{ulem}

\newcommand{\beginsupplement}{%
        \setcounter{table}{0}
        \renewcommand{\thetable}{S\arabic{table}}%
        \setcounter{figure}{0}
        \renewcommand{\thefigure}{S\arabic{figure}}%
     }

\begin{document}
\begin{CJK*}{UTF8}{gbsn}

\title{Influence of electronic entropy on Hellmann-Feynman forces in {\it ab initio} molecular dynamics with large temperature changes}

\author{Ming Geng (耿明)$^{1,2}$}
\author{Chris E. Mohn$^{1,2,3}$}
\email{chris.mohn@kjemi.uio.no}
\affiliation{
$^{1}$Centre for Planetary Habitability (PHAB), University of Oslo, N-0315 Oslo, Norway
}
\affiliation{
$^{2}$Centre for Earth Evolution and Dynamics (CEED), University of Oslo, N-0315 Oslo, Norway
}
\affiliation{
$^{3}$Department of Chemistry and Center for Materials Science and Nanotechnology, University 
of Oslo, Oslo 0371, Norway
}

\date{\today}

\begin{abstract}

The Z method is a popular atomistic simulation method for determining the melting temperature where a sequence of microcanonical molecular dynamics runs are carried out to target the lowest system energy where the solid always melts. Homogeneous melting at the limit of critical superheating, $T_{\text{h}}$, is accompanied by a drop in temperature as kinetic energy is converted to potential energy and the equilibrium melting temperature, $T_{\text{m}}$, can be calculated directly from the liquid state.
Implementation of the Z method interfaced with modern {\it ab initio} electronic structure packages use Hellmann-Feynman forces to propagate the ions in the microcanonical ensemble where the Mermin free energy plus the ionic kinetic energy is conserved. The electronic temperature, $T_{\text{el}}$, is therefore {\it kept fixed} along the trajectory which may introduce some spurious ion-electron interactions in molecular dynamics runs with large changes in temperatures such as often seen in homogeneous melting and freezing processes in the microcanonical ensemble. 
We estimate possible systematic errors in the calculated melting temperature to choice of $T_{\text{el}}$ for two main mantle components, the wide band-gap insulators SiO$_2$ and CaSiO$_3$ at high pressure. Comparison of the calculated melting temperature from runs where the $T_{\text{el}}$ =  $T_{\text{h}}$ and $T_{\text{el}}$ = $T_{\text{m}}$  - representing reasonable upper and lower boundaries respectively to choice of $T_{\text{el}}$ - shows that the difference in melting temperature is  200-300 K (3-5\% of the melting temperature) for our two test-systems.
Our results are in good agreement with previous large-size co-existence method and thermodynamic integration calculations, suggesting the CaSiO$_3$ and SiO$_2$ melts at around 6500 K (100 GPa) and 6000 K (160 GPa) respectively. The melting temperature decreases with increasing $T_{\text{el}}$ due to the increasing entropic stabilisation of the liquid and the systems melts typically about 3 times faster in molecular dynamics runs with $T_{\text{el}} = T_{\text{h}}$  compared to runs where $T_{\text{el}} = T_{\text{m}}$. 
A careful choice of electron temperature in Born-Oppenheimer molecular dynamic simulations where the ions are propagated using Hellmann-Feynamn forces with the Mermin free energy + the ionic kinetic energy being conserved, is therefore essential for the critical evaluation of the Z method and in particular at very high temperatures.

\end{abstract}
 
\maketitle
\end{CJK*}

\section{Introduction}

Melting and solidification processes are ubiquitous in condensed matter physics and in the evolution of terrestrial bodies including our own Earth. Triggered by extended defects, grain boundaries or open surfaces, equilibrium melting occurs spontaneously at $T_{\text{m}}$ when the free energy of the solid equals that of the liquid. Under certain conditions, however, a crystal can melt {\it homogeneously} at a much higher temperature, $T_{\text{h}}$, which represents the critical limit of superheating before melting is unavoidable~\cite{Belonoshko2006, Superheated2019}.
If a perfect periodic crystal in a molecular dynamics simulation is heated until $ T \approx T_{\text{h}}$, spontaneous fluctuations (nucleation precursors) will form transient defects and liquid nuclei which trigger an irreversible rapid nucleation growth and melting. Although homogeneous melting is rare in nature, shock-induced homogeneous melting has been demonstrated in a number of experiments (see for example~\cite {Russouw1985, Boyce1985, Cahn1986, Evans1986, Mo2018}). 

The Z method explores the link between homogeneous and equilibrium melting using molecular dynamics simulations and has been widely used to estimate melting temperatures for different materials ~\cite{Clathrate2011, kaoliniteZ2013, SiC2020, Sliver2021} often at high temperatures and pressure where experiments are hazardous or impractical~\cite{AlumZAIMD2009, Belonoshko2012, 1TPa2015, Osmium2015, Vanadium256GPa2020, Zirconium2022}. A number of molecular dynamics simulations are launched in the microcanonical (NVE) ensemble at different initial temperatures, $T_{\text{ini}}$, to target the lowest total energy, $E_{\text{h}}$, where the solid always melts. When the system melts at $E_\text{h}(T_{\text{h}}$), the temperature decreases to the equilibrium melting temperature while the latent heat of melting gradually converts into potential energy. If we assume a linear variation of energy with temperature we can establish a relationship between $T_{\text{h}}$ and $T_{\text{m}}$ from the entropy of melting~\cite{Belonoshko2006, Braithwaite2019}:
\begin{equation}
   \frac{T_{\text{h}}}{T_{\text{m}}} - 1 = \frac{\Delta S_{\text{m}}}{C_{\text{V}}}\label{eq1}
\end{equation}
where $C_{\text{V}}$ is the heat capacity at constant volume of the solid and $\Delta S_{\text{m}}$ is the entropy of melting. Eq.~\ref{eq1} can be approximated by ln2/3 assuming an ideal entropy of melting ($3k_\text{B}$ (per atom)) taken from a high-temperature limit of the Debye model and assuming that the heat capacity at constant volume is given by $k_\text{B}$ln2 (per atom) when $\Delta V_{\text{m}}/V \to 0 $ \cite{Stishov1973}.

Since the equilibrium melting temperature can be calculated at $T_{\text{h}}$  from a homogeneous melting triggered typically by small defects, quite small simulation boxes ($\sim$ 100 atoms) are in general sufficient to accurately calculate $T_{\text{h}}$ and hence $T_{\text{m}}$ from Eq.~\ref{eq1}~\cite{Davis2019}.
The use of small simulation boxes therefore makes the Z method a potentially attractive "low-cost" method for the accurate calculation of melting temperatures~\cite{Alfe2011, Gonzalez2016b, Davis2019}. By contrast, popular two-phase approaches to melting where the equilibrium melting process is mimicked by constructing a simulation box with a solid and a liquid phase in mechanical contact, often require large boxes to accommodate both solid and liquid phases and their interface~\cite{Hernandez2022, Brodholt2016}. 

In addition, since $T_m$ can be calculated from Eq. \ref{eq1}, there is no need for the explicit calculation of free energies which sometimes hamper the precision of thermodynamic integration~\cite{UP-TILD2009, TU-TILD2015} and 2PT methods~\cite{2PT2003}, especially when the solid and liquid free energy curves have very similar steepness near $T_{\text{m}}$~\cite{Melting2017, Hernandez2022}.

In spite of these advantages, recent studies have unveiled some artificial features that may hamper the Z method for the accurate calculation of melting temperatures, particularly under extreme conditions~\cite{Hernandez2022, Alfe2011}. Since the waiting time required for the solid to melt diverges in the limit where $T_{\text{ini}} \to T_\text{h}$, a waiting time analysis is often carried out at different $T_{\text{ini}} > T_\text{h}$ to estimate $T_{\text{m}}$ from extrapolation to that of "infinite" waiting-time~\cite{Alfe2011}. This analysis, however, may still require extensive statistics for the precise calculation of melting temperatures. 

Moreover, Born-Oppenheimer {\it ab initio} molecular dynamics (BOMD) is typically performed in the $NVE$ ensemble using Hellman-Faynman dynamics where the Fermi-Dirac electronic temperature is {\it kept fixed} in the MD simulation~\cite{ADV1992}. Although this implementation of the Z method ensures conserved dynamics~\cite{Wentzcovitch1992, Weinert1992}, large changes in temperature following melting and sometimes equilibration, may introduce systematic errors in $T_{\text{m}}$ since the electronic temperature is kept fixed~\cite{Hernandez2022}. That is, if a BOMD $NVE$ run with $E > E_{\text{h}}$ is launched with an $T_{\text{el}}$ {\it chosen near the melted liquid temperature} ($T_{\text{el}} \approx T_{\text{m}}$), then $T_{\text{el}}$ will be much lower than the temperature in the solid state (before melting). A too low electronic entropy may favour the stabilisation of the solid and prevent melting. This in turn will affect the estimated homogeneous melting temperature ($T_h$) and the "waiting time" for a solid to melt. The calculated equilibrium melting temperature may therefore be too high. On the other hand, if a BOMD run is launched with an electronic temperature {\it chosen near the solid temperature before melting} ($T_{\text{el}} \approx T_{\text{h}}$) a physically reasonable electronic-ionic interaction is ensured before melting. However, once the solid melts at constant volume, the temperature drops by 1 - $T_{\text{m}}/T_{\text{h}} \approx \text{ln}2/3$ and the electronic entropy will be much {\it higher} than the liquid temperature, which usually favours an entropic stabilisation of the liquid. Hence, the melting temperature may be too low. Therefore, since the choice of electronic entropy affect ion dynamics in processes undergoing large temperature drops, such as melting in the microcanonical ensemble, addressing the sensitivity in  $T_{\text{m}}$ to the choice of $T_{\text{el}}$ is crucial in order to benchmark the Z method for the accurate calculating of melting temperatures.

In this work, we thus investigate the role of electronic entropy on homogeneous melting for two main abundant mineral components in the Earth's interior CaSiO$_3$ and SiO$_2$ at high temperatures.  

Ca-perovskite is the third most abundant mineral in the lower mantle and a main component of basaltic lithologies constituting more than 20\% of recycled oceanic crust that is continuously being injected into the Earth's deep interior. 
A strong preferential partitioning of radioactive heat-producing elements into CaSiO$_3$, such as U and Th, as well as key geochemical tracers, suggests that CaSiO$_3$ is the main storage minerals for many of these minority elements~\cite{Corgne2005, Tateno2018}. Tracking the distribution of CaSiO$_3$ in the lowermost mantle is therefore essential to understand the evolution of the solid Earth which in turn requires the thermodynamic conditions of Ca-perovskite melting. Motivated by this, a  number of computational studies have calculated melting curves for pure CaSiO$_3$ to lowermost mantle conditions~\cite{Hernandez2022, Braithwaite2019}, but the agreement is not satisfactory. Here we attempt to contribute to tighten the constraints of CaSiO$_3$ melting at the lowermost mantle conditions.

Our second model system is SiO$_2$. The solid-liquid phase boundary of SiO$_2$ at ultrahigh pressure is critical to our understanding of not only the Earth's evolution but also the formation of many super-Earths~\cite{Das2020, Gonzalez2016a}. High-pressure silica melting may also play an important role in core dynamics, as it has been suggested that silica may have crystallized from a Si-saturated proto-core during a chemical exchange with a basal magma ocean~\cite{Hirose2017, Tronnes2019}. 
In spite of a number of simulations and experimental results reported in the literature, the SiO$_2$ melting curve remains poorly constrained at very high pressure. Here we will attempt to contribute to resolving some of these outstanding discrepancies.


\section{Theory}

The thermodynamic ensemble appropriate for the $Z$ method is the microcanonical ($NVE$) ensemble with the volume, $V$, number of species $N$ and the total system energy, $E$, kept fixed. The maximum energy along the solid branch of the isochore, $E_\text{h}$, is the same as the lowest energy along the liquid branch:
\begin{equation}
E_{\text{sol}}(V, T_{\text{h}}) = E_{\text{liq}}(V,T_{\text{m}}).\label{waitingtime} 
\end{equation}
To locate $E_{\text{sol}}(V, T_{\text{h}})$ and hence $E_{\text{liq}}(V, T_{\text{m}})$, a sequence of $NVE$ MD runs are carried out with different initial temperatures. Since the waiting time for the solid to melt diverges when $T \to T_\text{h}$, the calculated melting temperature will always represents an upper bound to the "true" melting temperature. 
To avoid extremely long MD runs in the vicinity of $T_\text{h}$, the melting temperature is calculated from an extrapolation of the distributions of waiting times using       
\begin{equation}
  \langle \tau \rangle^{-1/2} = A(T_{\text{liq}} - T_{\text{m}})
  \label{Eqwait}
\end{equation}
where "A" is a parameter, $\tau$ is the waiting time for a solid to melt at a given total energy and $T_{\text{liq}}$ is the liquid temperature of the system after melting. Since $T_{\text{liq}} = T_{\text{m}}$, when $E = E_{\text{h}}$, 
the melting temperature can be found at infinite waiting time i.e. at the point of intersection where $\langle \tau \rangle^{-1/2} = 0$.  

We use {\it ab initio} Born-Oppenheimer MD to propagate the ions where the electronic energy is minimized at each step along the trajectory. Note that the usual Hellmann-Feynman forces do not conserve the total system energy, $E = U + K$ where $U$ is the internal DFT energy and $K$ is the kinetic energy of the ions. This is because the energy functional is non-variational with respect to changes in partial orbital occupancies along the MD trajectory.

To avoid additional contributions to the Hellmann-Feynman forces due to the variation in the band occupancies along the ionic trajectory the quantity  "$K + \Omega$", rather than the total energy, is conserved in the MD run where $\Omega$ is the Mermin free-energy~\cite{Mermin1965, Wentzcovitch1992, Weinert1992} 
\begin{equation}
    \Omega = U-T_{\text{el}}S_{\text{el}}
    \label{mermin}
 \end{equation} 
with 
\begin{equation}
    S_{\text{el}}=-k_{\text{B}}\sum _i[f_i \text{ln}{f_i}+(1-f_i)\text{ln}(1-f_i)].
\end{equation}
$S_{\text{el}}$ is the electronic entropy and $f_i$ is the electron occupancy of band $i$ calculated using Fermi-Dirac statistics: 
\begin{equation}
   f_i=F(\dfrac{\epsilon_i-E_{\text{fermi}}}{\sigma}),  
   \sum_{0}^{N_i}f_i=N 
\end{equation}
where $F$ is the usual (Fermi-Dirac) smearing function, $\epsilon_i$ is the eigenvalue of band $i$, $E_{\text{fermi}}$ is the Fermi level and $\sigma$ = k$_\text{B}/T$ is the smearing broadening.  

Ionic dynamics may be sensitive to the choice of electronic temperature when the ions are propagated using Hellmann-Feynman forces, in particular when the temperature changes are large such as during equilibration and melting in the $NVE$ ensemble. Electronic temperature may therefore affect the melting processes for semiconductors and insulators even though the fractional occupancies of the conduction bands, remain small during these temperature changes.

We can estimate the sensitivity in the calculated $T_{\text{m}}$ to choices of $T_{el}$ by comparison of the melting temperature calculated using $T_{\text{el}} \approx T_{\text{h}}$ with that calculated $T_{\text{el}} \approx T_{\text{m}}$. These choices of $T_{el}$ provide reasonable upper and lower bounds to the calculated melting temperature in the microcanonical ensemble when the Hellmann-Feynman forces are used for the ionic propagation with $K + \Omega$ being conserved. 

If changes in $T_{\text{el}}$ can not be ignored, the usual conservation law including the Mermin functional in the form of Eq.~\ref{mermin} must be replaced, but the forces will then include contributions arising due to changes in the partial orbital occupancies along the microcanonical Born-Oppenheimer trajectory. It is important to note, however, that the time evolution of orbital occupancies due to large temperature changes may be only correctly described by the time-dependent Schr\"odinger equation.

A possible strategy to calculate $T_{\text{m}}$  using the Z method is to adjust $T_{\text{el}}$ along the ionic trajectory to match the average (ionic) temperature in the previous $N$ time-steps~\cite{Hernandez2022}. Although this "update scheme" does not ensure conserved dynamics, the average ensemble temperature before and after adjustment is in general expected to be small. In this approach, the time-dependent Schr\"ordinger equation is thus approximated by a sequence of BOMD NVE simulations.

\section{Computational details}

All BOMD simulations are performed with the Vienna {\it ab initio}  simulation package (VASP) \cite{VASP93, VASP94}, using the projector augmented wave (PAW) method \cite{PAW94, PAW99}. We use the generalized gradient approximation (GGA) where the exchange-correlation contribution to the energy is parameterized using the PBE~\cite{PBE96} functional for SiO$_2$ and the AMO5 functional~\cite{Armiento2005} for CaSiO$_3$. The electronic configurations were: [He]2$s^2$2$p^4$ for O, [Ne]$3s^23p^2$ for Si and [He]$3s^3p^64^s2$ for Ca. The energy cutoff for the plane wave was 700 eV for SiO$_2$ and somewhat lower, 500 eV, for CaSiO$_3$ to compare directly with previous CaSiO$_3$  DFT studies~\cite{Hernandez2022, Braithwaite2019}.

In all runs, the atoms were placed at their ideal crystallographic sites i.e. the $1b$,  $1a$ and $3d$ positions for Ca, Si and O atoms respectively of the $Pm3m$  space group (CaSiO$_3$). SiO$_2$ MD runs were started from the ideal cubic pyrite-type structure ($Pa\bar{3}$) optimized to target an equilibrium pressure $\sim$ 160 GPa. This is probably slightly below the stability field of the pyrite structured SiO$_2$ near the melting curve~\cite{Das2020}, but in order to compare directly with results in Ref.~\cite{Gonzalez2016a, Gonzalez2016b} we use pyrite rather than seifertite. The estimated melting point when pyrite is used as the crystal structure is only slightly lower compared to that found using seifertite.~\cite{Gonzalez2016a, Gonzalez2016b}.
For CaSiO$_3$ we use a cubic 135 atoms simulation box which is the same as that used in previous computational studies of CaSiO$_3$ melting~\cite{Braithwaite2019, Hernandez2022} allowing for a direct comparison with these studies. For SiO$_2$ the simulation box  contained 96 atoms.


Melting simulations are carried out in the $NVE$ ensemble with a timestep of 0.5 fs for SiO$_2$ and 1 ps for CaSiO$_3$. The smaller timestep for SiO$_2$ was chosen to minimize the energy fluctuation. 
All runs used in the calculation of the waiting time were carried out until melting plus an additional 5-20 ps to calculate the average liquid temperature.

The waiting time analysis were performed based on between 8 and 20 different simulations at a given (E,V) where, in each run, the forces were taken from a Maxwell-Boltzmann distribution. Close to the equilibrium melting temperature we performed typically around 20 MD runs at a given ($E,V$) to ensure that sufficient statistics were collected in order to calculate $\langle \tau \rangle$ using Eq.~\ref{Eqwait}. Plots of the convergence of the estimated melting temperature with number of configurations are shown in supplementary information. This analysis shows that about 8-10 MD runs, for a given initial temperature, is sufficient to converge the melting temperature to less than 100 K. All MD calculations launched below $E_{\text{h}}$, lasted for at least 10 ps and close to $T_{\text{h}}$ the MD simulations typically ran for more than 100 ps. 

For SiO$_2$, the $T_{\text{el}}$ are 6000 K, 7000 K and 8000 K where 6000 K is expected, after test-calculations, to lie close to $T_{\text{m}}$ whereas 8000 K will lie close to $T_{\text{h}}$. Similarly, for CaSiO$_3$ the waiting time analysis was carried out at $T_{\text{el}}$ = 6500 K and 9000 K which are expected to be close to $T_{\text{m}}$  and $T_{\text{h}}$  respectively.
To simulate melting with negligible contribution from the electronic entropy we used a Gaussian scheme~\cite{ADV1992} with a very low value of the smearing parameter (i.e. $\sigma_{\text{Gaussian}}$ = 0.03 eV).
The Gaussian smearing method is better designed to avoid instabilities arising from fluctuations in orbital occupancies at low values of $\sigma$ (low temperatures) which often hamper the Fermi-Dirac method during energy minimizations. The Gaussian smearing has the functional form $\frac{1}{2}(1-\text{erf}\dfrac{\epsilon-\mu}{\sigma})$ and the link between the two schemes is given by the ratios of the full width at half maximum, FWHM, as:
\begin{equation}
   \frac{\text{FWHM}_{\text{FermiDirac}}}{\text{FWHM}_{\text{Gaussian}}}= \dfrac{\text{cosh}^{-1}(\sqrt{2})}{\sqrt{\text{ln}2}}.
\end{equation} 

\section{Results and Discussion}

\subsection{Influence of electronic entropy on ionic dynamics}

\begin{figure*}
    \includegraphics[width=\textwidth]{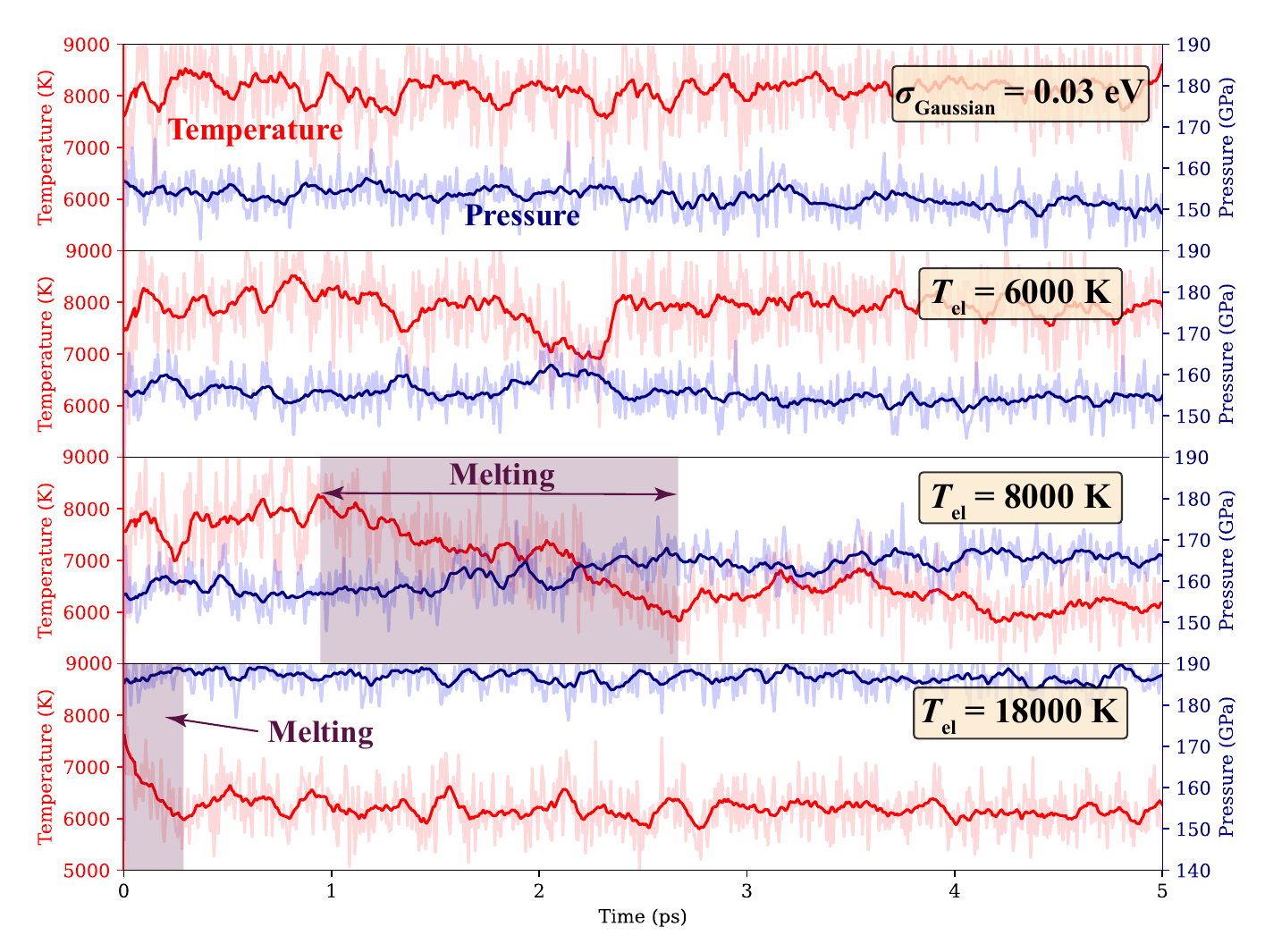}
    \caption{Temperature (red) and pressure (blue) evolutions along snapshots of {\it ab initio} MD trajectories for SiO$_2$ with different $T_{\text{el}}$. The thin lines show dumps every time-step whereas the thick lines are averaged properties over the previous 100 time-steps. All MD runs are carried out with the same initial temperature ($T_{\text{ini}}=18000$ K) where the atoms are distributed at their equilibrium lattice positions before we launch the MD run. A Gaussian smearing broadening scheme with $\sigma_{\text{Gaussian}}$ = 0.03 eV is used to represent simulations with a negligible contribution to electronic entropy on the ionic dynamics. The shaded area shows the melting process.}
    \label{fig1}
\end{figure*}

In Fig.~\ref{fig1} we illustrate the sensitivity to changes in electronic temperature on the melting dynamics for silica where the initial ionic temperatures are {\it the same} in all runs.
In the extreme case where the electronic temperature is the same as the initial temperature in the MD runs ($T_{\text{el}}  = T_{\text{ini}} $ = 18000 K), SiO$_2$ always melts rapidly and instantaneously in less than $ 0.5$ ps. On the contrary, in runs with negligible contribution from electronic entropy (i.e. with $\sigma_{\text{Gaussian}}$ = 0.03 eV), melting is rare and we observed only one incidence of melt-nucleation in all our 20 runs which lasted 10 ps each.

In simulations with "intermediate" $T_{\text{el}}$, close to either the homogeneous or the equilibrium melting temperatures (i.e. with  $T_{\text{el}} = 8000$ K or $T_{\text{el}}=6000$ K respectively), the system with $T_{\text{el}} = 6000$ K melted markedly slower compared to those with $T_{\text{el}} = 8000$ K and the waiting time was much longer. We found that the average waiting time was 2.7 ps when $T_{\text{el}}$ = 8000 K $\approx T_{\text{h}}$ and markedly longer by a factor of about 3 when  $T_{\text{el}} = 6000 \approx T_{\text{m}}$. See also the discussion about distributions of the waiting time in the supplementary information. Interestingly, when $T_{\text{el}} = 6000$ K, we observe that the temperature sometimes drops markedly indicating possibly the formation of melt nuclei, but the system quickly reverted to its original (solid) state. This is seen as a small "bump" in the temperature/pressure evolution in  Fig.~\ref{fig1} at about 2.0-2.5 ps. Solid-liquid "oscillations" are often seen in runs where the energy is close to the target homogeneous melting energy using small simulation boxes with large temperature fluctuations~\cite{Alfe2011} and will be discussed more in the supplementary information. 

\begin{figure*}
   \includegraphics[width=0.90\textwidth]{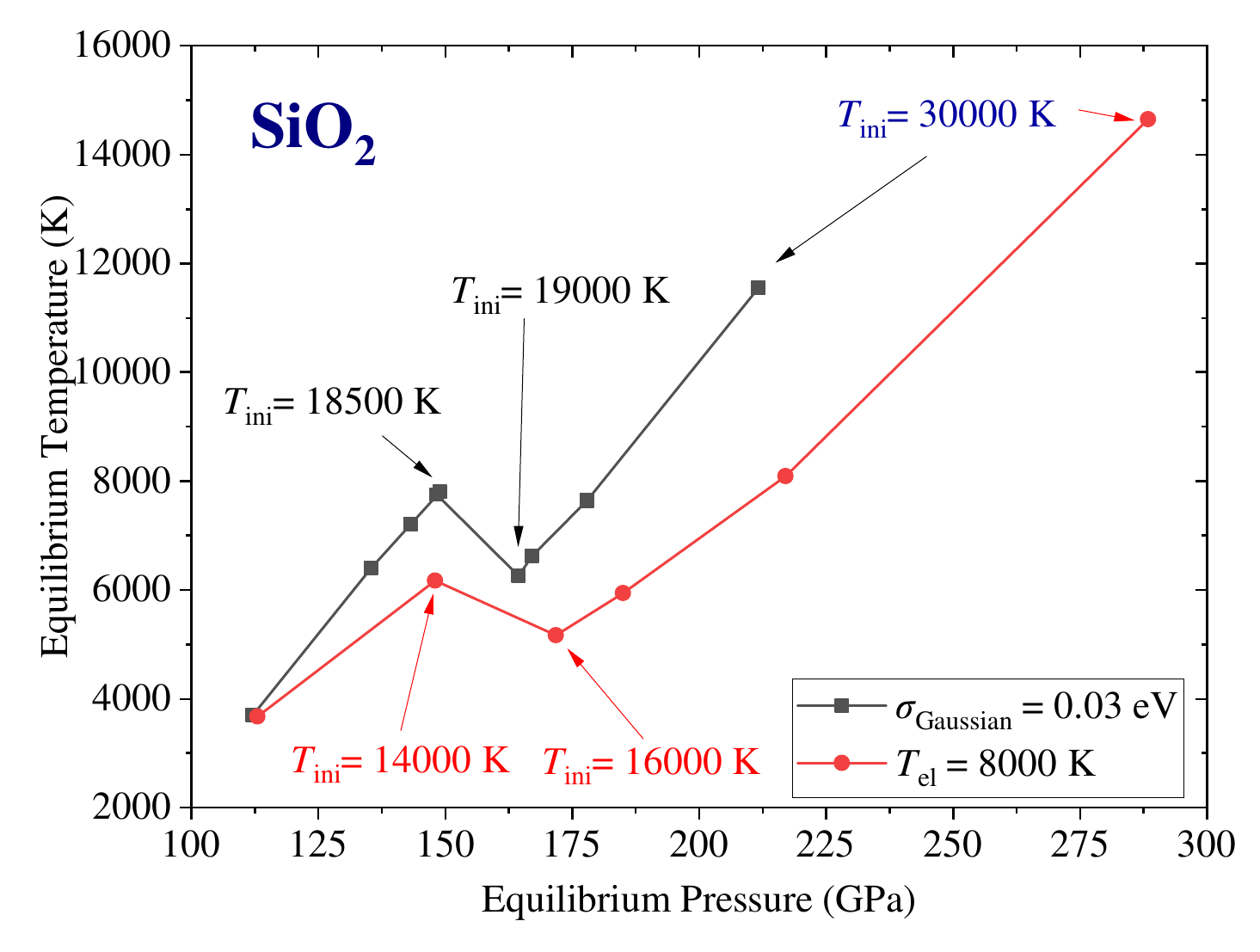}

    \caption{Two isochores with different electronic temperatures. The  black curve is runs where $\sigma _{\text{Gaussian}}$ = 0.03 eV  and the red curve is runs where $T_{\text{el}} = 8000 \approx T_{\text{m}}$. The initial temperatures are in the range 8000 K to 30000 K with the ions initially placed at their ideal lattice positions before we launched the MD simulations which ran for 20 ps each.}
    \label{fig2}
\end{figure*}

The influence of electronic entropy on properties is also seen in Fig.~\ref{fig2} where two isochores with different electronic temperatures are drawn. Here, the initial ionic temperature in the MD runs is systematically increased from 8000 K to 30000 K. The isochore with an high electronic temperature i.e. close to $T_{\text{h}}$ (i.e. with $T_{\text{el}} = 8000$ K) deviates strongly from the one with negligible contributions from electronic entropy (i.e. $\sigma _{\text{Gaussian}}$ = 0.03 eV).



\subsection{Choice of electronic temperature in {\it ab initio} MD runs}

\begin{figure*}
    \includegraphics[width=0.94\textwidth]{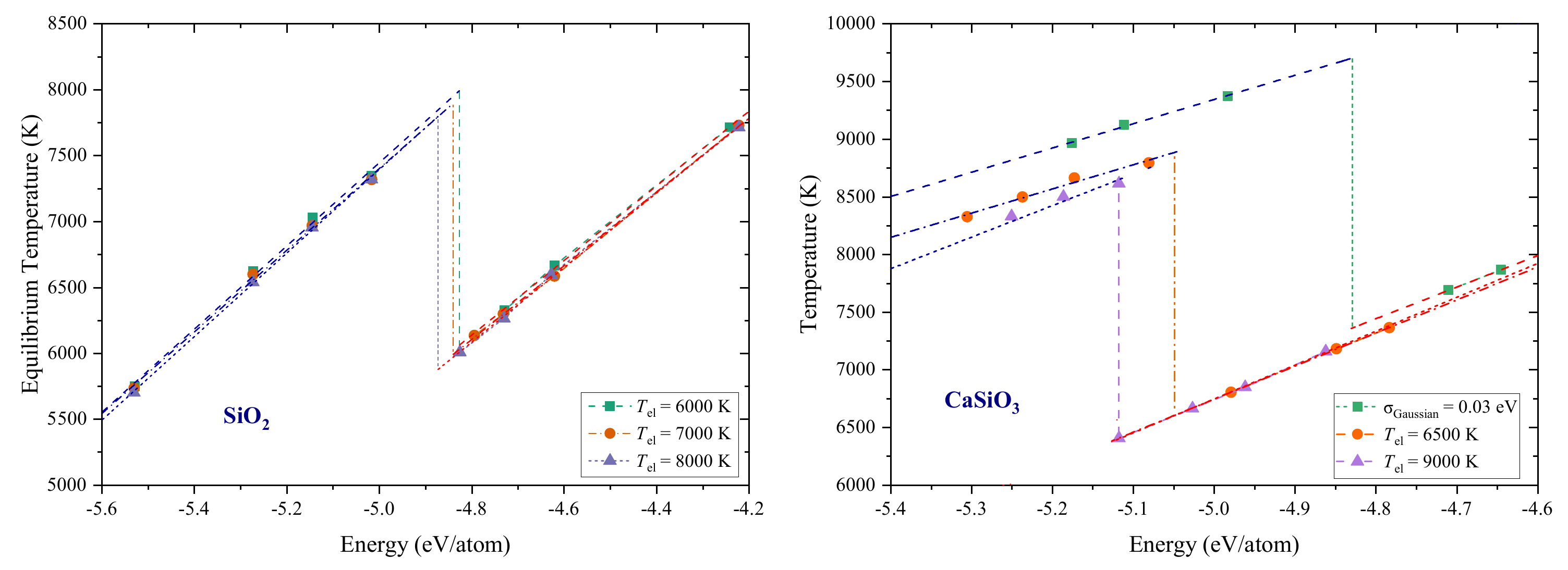}
    \caption{Solid (blue) and liquid (red) branches in MD-$NVE$ runs for SiO$_2$ (left) and CaSiO$_3$ (right). For SiO$_2$, $T_{\text{el}}$ = 6000 K, 7000 K and 8000 K whereas for CaSiO$_3$, $T_\text{el}$ are 6500 K and 9000 K. In addition, we plot the result for CaSiO$_3$ with a Gaussian smearing scheme using $\sigma_{\text{Gaussian}} = 0.03$ eV. The homogeneous melting temperatures are calculated from an intersection of a linear extrapolation of the solid branch runs and a vertical line drawn from the equilibrium melting point (calculated using Eq. 3). The resulting homogeneous melting temperatures may therefore represent an upper bound to the "true" homogeneous melting temperature since the slope of the solid branch typically decreases near $T_{\text{h}}$~\cite{Davis2019,Hernandez2022}. Values of  $T_{\text{h}}$ are reported in Table \ref{table}.}
    \label{fig3}
\end{figure*}
\begin{figure*}
    \includegraphics[width=0.95\textwidth]{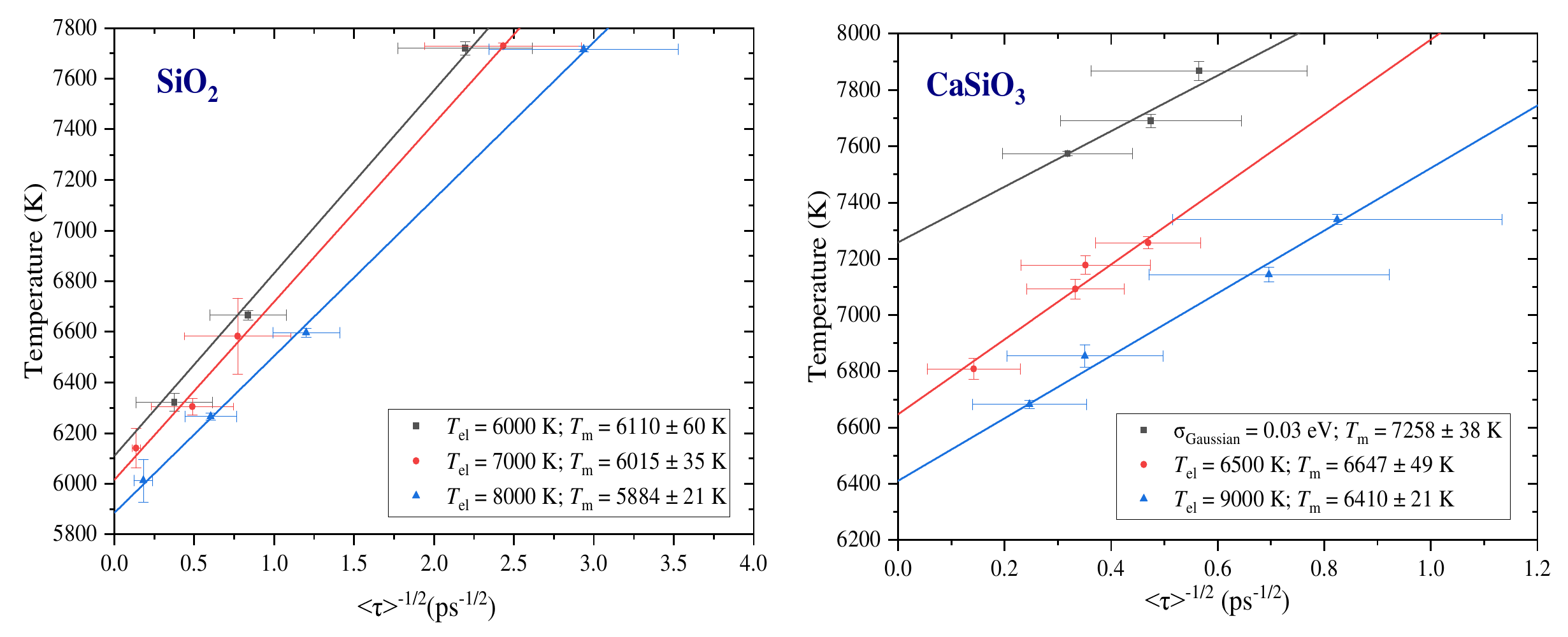}
    \caption{The calculated melting temperature using Eq.~\ref{Eqwait} for different choices of electronic temperatures. For SiO$_2$, $T_{\text{el}}$ = 6000 K, 7000 K and 8000 K whereas for CaSiO$_3$, $T_\text{el}$ are 6500 K and 9000 K. In addition, we plot the waiting time for CaSiO$_3$ using a Gaussian smearing scheme with $\sigma_{\text{Gaussian}} = 0.03$ eV. The resulting estimated equilibrium melting temperatures are reported in Table 1~\ref{table} along with the standard error. The horizontal and vertical error bars reported are the mean errors of the waiting time and temperatures respectively.}\label{fig4}
\end{figure*}

The sensitivity in melting temperature to choice of $T_{\text{el}}$ illustrated above for SiO$_2$ and discussed in the supplementary information suggests that BOMD runs in the $NVE$ ensemble where the ions are propagated using the Hellmann-Feynman forces with $\Omega  + K$ being conserved may introduce some errors due to changes in electron-ion interactions following large drops in temperature $ \approx$ ln2/3 $\times T_{\text{h}}$. This is seen in the Z plots and the waiting-time analysis in Fig.~\ref{fig3} and~\ref{fig4}  as well as in Table~\ref{table} for both SiO$_2$ and CaSiO$_3$. When $T_{\text{el}}$ is kept fixed at some value near $T_{\text{m}}$, the calculated equilibrium melting temperature will be markedly higher compared to that if $T_{\text{el}}$ is close to the homogeneous melting temperature.
The calculated melting temperatures with  $T_{\text{el}} \approx T_{\text{m}}$ at around 100 GPa (CaSiO$_3$) and 150 GPa (SiO$_2$) are about 300 K and 200 K higher respectively than those calculated with $T_{\text{el}}  \approx T_{\text{h}}$. This corresponds roughly to 10\% of the temperature drop accompanying melting. We expect that these absolute errors increase with increasing melting temperature and pressure since the temperature drop accompanying melting increases with increasing $T_{\text{h}}$.

A key question is therefore: what is the best choice of electronic temperature in the MD runs to minimize the errors in the calculated $T_{\text{m}}$ when the ions are propagated using Hellmann-Feynmann forces with $\Omega + K$ being conserved?
If we choose an electronic temperature very close to $T_{\text{h}}$ in MD runs with $ E = E_{\text{h}}$, the electronic temperature is very close to the (ensemble) average temperature before the system eventually melts. Runs with $T_{\text{el}} \approx T_{\text{h}}$ therefore enables the accurate calculation of $T_{\text{h}}$  as well as the waiting time and implies that MD runs where $T_{\text{el}} < T_{\text{h}} $  (if, say  $T_{\text{el}} = T_{\text{m}}$) give too high homogeneous melting temperatures. 
Although $T_{\text{el}} \approx T_{\text{h}}$ (with $E \approx E_{\text{h}}$) enables the accurate calculation of $T_{\text{h}}$, the melting temperature may be severely underestimated. This is because $T_{\text{el}}$ is much larger than the liquid (ionic) temperature which, in general, favours an entropic stabilization of the liquid over the solid and hence the calculated melting temperature will be too low.  
If we rather chose $T_{\text{el}} \approx T_{\text{m}}$ -  which is the lowest temperature on the isochore and therefore represents a reasonable lower bound to choice of $T_{\text{el}}$ -  the homogeneous melting temperature and possibly the equilibrium melting temperature will be {\it overestimated}. Calculations where $T_{\text{el}}$ is chosen to target either $T_{\text{m}}$ or T$_{\text{h}}$, therefore, provide reasonable {\it upper and lower bounds} respectively to the true melting temperature. 

\subsection{CaSiO$_3$ melting}

As shown in Table \ref{table}, our melting temperatures are in very good agreement with a previous Z method study~\cite{Hernandez2022} and also in excellent agreement with those from thermodynamic integration and two-phase calculations~\cite{Hernandez2022}. The inclusion of electronic entropy is essential for the accurate calculation of melting temperature for CaSiO$_3$, in agreement with that found for other wide band gap insulators such as MgO ~\cite{Alfe2005}. That is, the calculated $T_{\text{h}}$ and $T_{\text{m}}$ without contribution from the electronic entropy are about 1000 K and 650 K higher respectively than those calculated using $T_{\text{el}}$ = 6500 K. We find that that the melting temperature is around 6300 K (with $T_{\text{el}}$ = 9000 K) and 6600 K (with $T_{\text{el}}$ = 6500 K). 
As discussed above, the discrepancy between the calculated melting temperature using $T_{\text{el}}$ = 6500 K with that calculated using $T_{\text{el}}$ = 9000 K is non-negligible for the accurate calculation of melting temperature. This places some constraints on the accuracy of the Z method interfaced with BOMD where $\Omega$ + K is conserved and the forces are propagated using the Hellmann-Feynman forces in the $NVE$ ensemble.

Choice of electronic entropy could therefore possibly explain the large discrepancy of more than 1000 K between our melting point and that of a recent Z method study~\cite{Braithwaite2019} which was carried out at the same thermodynamic conditions as here. We use the same exchange-correlation functional to DFT as that employed in Ref.~\cite{Braithwaite2019}, but are unable to reproduce their melting point unless we set  $T_{\text{el}}$ = $T_{\text{ini}}$. Ref.~\cite{Braithwaite2019}, however, do not report values of  $T_{\text{el}}$ so no firm conclusions can be drawn.



\begin{table*}[]
\setlength{\tabcolsep}{3pt}
\renewcommand{\arraystretch}{1.2}
\begin{tabular}{ccccccc}
\hline
\hline
\textbf{System} &
  \textbf{Method} &
  \textbf{$T_{el}$ (K)} &
  \textbf{P (GPa)} &
  \textbf{$T_h$ (K)} &
  \textbf{$T_m$ (K)} &
  \textbf{Ref.} \\ \hline
\multicolumn{1}{c|}{\multirow{9}{*}{\textbf{CaSiO$_3$}}} &
  \multicolumn{1}{c|}{Z-Method} & \multicolumn{1}{c|}{9000} &  \multicolumn{1}{c|}{103.7$\pm$2.6} & \multicolumn{1}{c|}{8873$\pm$42} &   \multicolumn{1}{c|}{6410$\pm$21} &  This work \\
\multicolumn{1}{c|}{} &
  \multicolumn{1}{c|}{Z-Method} &
  \multicolumn{1}{c|}{6500} &
  \multicolumn{1}{c|}{103.0$\pm$2.5} &
  \multicolumn{1}{c|}{8652$\pm$25} &
  \multicolumn{1}{c|}{6647$\pm$49} &
  This work \\
\multicolumn{1}{c|}{} &
  \multicolumn{1}{c|}{Z-Method} &
  \multicolumn{1}{c|}{$\sim$0} &
  \multicolumn{1}{c|}{105.2$\pm$3.0} &
  \multicolumn{1}{c|}{9605$\pm$11} &
  \multicolumn{1}{c|}{7258$\pm$38} &
  This work \\
\multicolumn{1}{c|}{} &
  \multicolumn{1}{c|}{Z-Method} &
  \multicolumn{1}{c|}{adjust$^a$} &
  \multicolumn{1}{c|}{102.6$\pm$2.7} &
  \multicolumn{1}{c|}{8506} &
  \multicolumn{1}{c|}{6517} &
  This work \\
\multicolumn{1}{c|}{} &
  \multicolumn{1}{c|}{Z-Method} &
  \multicolumn{1}{c|}{adjust$^b$} &
  \multicolumn{1}{c|}{105$\pm$3.3} &
  \multicolumn{1}{c|}{8806} &
  \multicolumn{1}{c|}{6493} &
  Hernandez {\it et al.} 2022~\cite{Hernandez2022} \\
\multicolumn{1}{c|}{} &
  \multicolumn{1}{c|}{Z-Method} &
  \multicolumn{1}{c|}{not reported} &
  \multicolumn{1}{c|}{103.0$\pm$0.2} &
  \multicolumn{1}{c|}{7120} &
  \multicolumn{1}{c|}{5200} &
  Braithwaite {\it et al.} 2019~\cite{Braithwaite2019} \\
\multicolumn{1}{c|}{} &
  \multicolumn{1}{c|}{large-size co-existence} &
  \multicolumn{1}{c|}{} &
  \multicolumn{1}{c|}{$\approx$103} &
  \multicolumn{1}{c|}{} &
  \multicolumn{1}{c|}{6582} &
  Hernandez {\it et al.} 2022~\cite{Hernandez2022}\\
\multicolumn{1}{c|}{} &
  \multicolumn{1}{c|}{Thermodynamic Integration} &
  \multicolumn{1}{c|}{} &
  \multicolumn{1}{c|}{$\approx$103} &
  \multicolumn{1}{c|}{} &
  \multicolumn{1}{c|}{6433} &
  Hernandez {\it et al.} 2022~\cite{Hernandez2022} \\
\multicolumn{1}{c|}{} &
  \multicolumn{1}{c|}{2-Phase thermodynamics} &
  \multicolumn{1}{c|}{} &
  \multicolumn{1}{c|}{$\approx$103} &
  \multicolumn{1}{c|}{} &
  \multicolumn{1}{c|}{5420} &
  Hernandez {\it et al.} 2022~\cite{Hernandez2022} \\ \hline
\multicolumn{1}{c|}{\multirow{9}{*}{\textbf{SiO$_2$}}}&
  \multicolumn{1}{c|}{Z-Method} &
  \multicolumn{1}{c|}{6000} &
  \multicolumn{1}{c|}{164$\pm$4.3} &
  \multicolumn{1}{c|}{8058$\pm$39} &
  \multicolumn{1}{c|}{6110$\pm$60} &
  This work \\
\multicolumn{1}{c|}{} &
  \multicolumn{1}{c|}{Z-Method} &
  \multicolumn{1}{c|}{7000} &
  \multicolumn{1}{c|}{164$\pm$4.3} &
  \multicolumn{1}{c|}{7899$\pm$30} &
  \multicolumn{1}{c|}{6015$\pm$35} &
  This work \\
\multicolumn{1}{c|}{} &
  \multicolumn{1}{c|}{Z-Method} &
  \multicolumn{1}{c|}{8000} &
  \multicolumn{1}{c|}{166$\pm$4.3} &
  \multicolumn{1}{c|}{7789$\pm$17} &
  \multicolumn{1}{c|}{5884$\pm$21} &
  This work \\
\multicolumn{1}{c|}{} &
  \multicolumn{1}{c|}{Z-Method} &
  \multicolumn{1}{c|}{adjust$^a$} &
  \multicolumn{1}{c|}{169$\pm$4.3} &
  \multicolumn{1}{c|}{} &
  \multicolumn{1}{c|}{6044} &
  This work \\
\multicolumn{1}{c|}{}  &
  \multicolumn{1}{c|}{Coexist} &
  \multicolumn{1}{c|}{} &
  \multicolumn{1}{c|}{153.8} &
  \multicolumn{1}{c|}{} &
  \multicolumn{1}{c|}{5990} &
  Benlonoshko {\it et al.} 1995~\cite{Belonoshko1995} \\
\multicolumn{1}{c|}{} &
  \multicolumn{1}{c|}{Coexist} &
  \multicolumn{1}{c|}{} &
  \multicolumn{1}{c|}{157.6} &
  \multicolumn{1}{c|}{} &
  \multicolumn{1}{c|}{5986} &
  Usui {\it et al.} 2010~\cite{Usui2010}\\
\multicolumn{1}{c|}{} &
  \multicolumn{1}{c|}{Shock Experiment} &
  \multicolumn{1}{c|}{} &
  \multicolumn{1}{c|}{157.0} &
  \multicolumn{1}{c|}{} &
  \multicolumn{1}{c|}{5543} &
  Millot {\it et al.} 2015~\cite{Millot2015} \\
\multicolumn{1}{c|}{} &
  \multicolumn{1}{c|}{Z-Method} &
  \multicolumn{1}{c|}{not reported} &
  \multicolumn{1}{c|}{132.3} &
  \multicolumn{1}{c|}{} &
  \multicolumn{1}{c|}{5852} &
  Gonz\'alez-Cataldo {\it et al.} 2016~\cite{Gonzalez2016a} \\
\multicolumn{1}{c|}{} &
  \multicolumn{1}{c|}{DAC Experiment} &
  \multicolumn{1}{c|}{} &
  \multicolumn{1}{c|}{117} &
  \multicolumn{1}{c|}{} &
  \multicolumn{1}{c|}{$\approx$6200} &
  Andrault {\it et al.} 2022~\cite{Andrault2022} \\
 \hline \hline
\end{tabular}
\caption{Calculated melting temperature for CaSiO$_3$ and SiO$_2$ using the Z method together with previous values reported in the literature at similar pressure (i.e. about 103 GPa for CaSiO$_3$ and 160 GPa for SiO$_2$). 
In the simulation where $T_{\text{el}} \sim 0$ we used Gaussian smearing, $\sigma _{\text{gauss}}$ = 0.03 eV, for the partial occupancies of the one-electron orbitals. In the simulation from Ref.~\cite{Hernandez2022} where $T_{\text{el}}$ is labelled as "adjust$^b$", the Fermi-Dirac smearing was updated about every 1 ps along the MD trajectory to match the average temperature in the previous $\sim$ 1 ps,  as discussed in Ref.~\cite{Hernandez2022}. In "adjust$^a$" we update the electronic entropy only once along the MD trajectory after about 25 ps of propagation in the liquid state. 
}\label{table}
\end{table*}


\subsection{SiO$_2$ melting}
There are not many experimental studies of SiO$_2$ that report melting temperatures to very high pressure. Results from a recent high-pressure experimental study~\cite{Millot2015} was fitted to an equilibrium melting curve to about 500 GPa ($T_\text{m}(P)=1968.5+307.8\times P^{0.485}$) suggesting that  SiO$_2$ melts at around 5540 K at 157 GPa. This is slightly lower than that calculated from a Z method simulation \cite{Gonzalez2016a},  where the melting temperature was estimated to be 5850 K at 132 GPa.  By contrast, a recent Diamond Anvil Cell (DAC) experiment \cite{Andrault2022, Andrault2020},  suggests a markedly higher melting curve compared to those mentioned above and the Clapeyron slope is also much steeper in the pressure region 120-150 GPa compared to that reported by e.g. Millot {\it et al}~\cite{Millot2015}. 
The melting curves from two molecular dynamics simulations \cite{Belonoshko1995, Usui2010} using two-phase co-existing methods are in overall good agreement with that from the DAC study~\cite{Andrault2022, Andrault2020}, but without the rapid change in the Clapeyron slope seen in the DAC experiment at around 120 GPa.

Our calculated melting temperatures reported in Table ~\ref{table} are in overall good agreement with previous computational predictions at similar pressures~\cite{Belonoshko1995, Usui2010}.
The calculated melting temperature reported using $T_{\text{el}}$ = 8000 K, for example, is 5776 K. This is only slightly lower compared to those of Refs.~\cite{Belonoshko1995} and ~\cite{Usui2010}  which are 5990 K and 5986 K respectively. 
Of note is that the good agreement with that reported by Usui and Tschuchia~\cite{Usui2010} may be fortuitous because a very small two-phase simulation box containing only 48 atoms for each phase was used in Ref.~\cite{Usui2010}. Such a small box has a boundary that is of similar size as the solid and liquid portions, and many runs are needed at a given (E,V) to precisely determine the melting points~\cite{Hong2016}. Our result also suggests that the predicted equilibrium melting curve reported from a shock experiment study~\cite{Millot2015} may be too low since it is assumed that stishovite is able to crystallize at the time-scale of the experiment and the melting curve is therefore drawn at the bottom of the liquid branch of the Hugoniot. If, however, stishovite is unable to crystallize at the time-scale of the shock experiment (i.e. within a few nanoseconds), the melting temperature reported in Ref.~\cite{Millot2015} may be underestimated as suggested by ~\cite{Shen2016}. This interpretation is consistent with our calculated melting point.


As discussed above, the calculated melting temperature using $T_{\text{el}}$ = 8000 K is slightly less than 200 K lower than that calculated using $T_{\text{el}}\approx T_{\text{h}} \approx 6000$ K.  This difference in the calculated melting temperature is of similar size compared to that found for CaSiO$_3$ at similar conditions and confirms that the Z method may be hampered by some artificial feature for the accurate calculations of melting temperature with the Mermin free energy + ionic kinetic energy being conserved along the Born-Oppenheimer MD-NVE trajectory with $T_{\text{el}}$ kept fixed.

\subsection{Sensitivity to changes in electronic entropy on the waiting time analysis}

The waiting time for the solid to melt is correctly described if we chose $T_{\text{el}} \approx \langle T \rangle_{\text{sol}}$  after equilibration (i.e. if, for example, $T_{\text{el}} = T_{\text{h}}$ when $E = E_{\text{h}}$). However, since the melting temperature is in general underestimated with  $T_{\text{el}} = T_{\text{h}}$, interpolation to infinite waiting time using Eq.~\ref{Eqwait} gives a too low equilibrium melting temperature when $\langle \tau \rangle $ is extrapolated to infinite waiting time.
However, if we use a lower $T_{\text{el}}$ (i.e. if we chose $T_{\text{el}} = \langle T \rangle _{\text{liq}}$) the waiting time for the system to melt will be too slow. This implies that  $\langle \tau \rangle ^{-1/2}$ should be shifted to lower temperatures indicating that the melting temperature calculated from the intersection $\langle \tau \rangle^{-1/2}= 0$ will be overestimated. The estimated melting temperatures from a waiting time analysis with $T_{\text{el}} \approx T_{\text{m}}$ and $T_{\text{el}} \approx T_{\text{h}}$, therefore, provide reasonable upper and lower bounds respectively to the "true" equilibrium melting temperature. These differences ($\sim$ 200-300 K) are much larger than the standard errors from the waiting time analysis (reported in Table \ref{table}) which, in our case, are always less than 60 K. Convergence plots of the calculated melting temperatures using Eq. 3 with number of MD runs show that only a few tens of MD calculation are needed for the accurate calculation of melting temperatures (see supplementary information).

\begin{figure*}
    \includegraphics[width=0.9\textwidth]{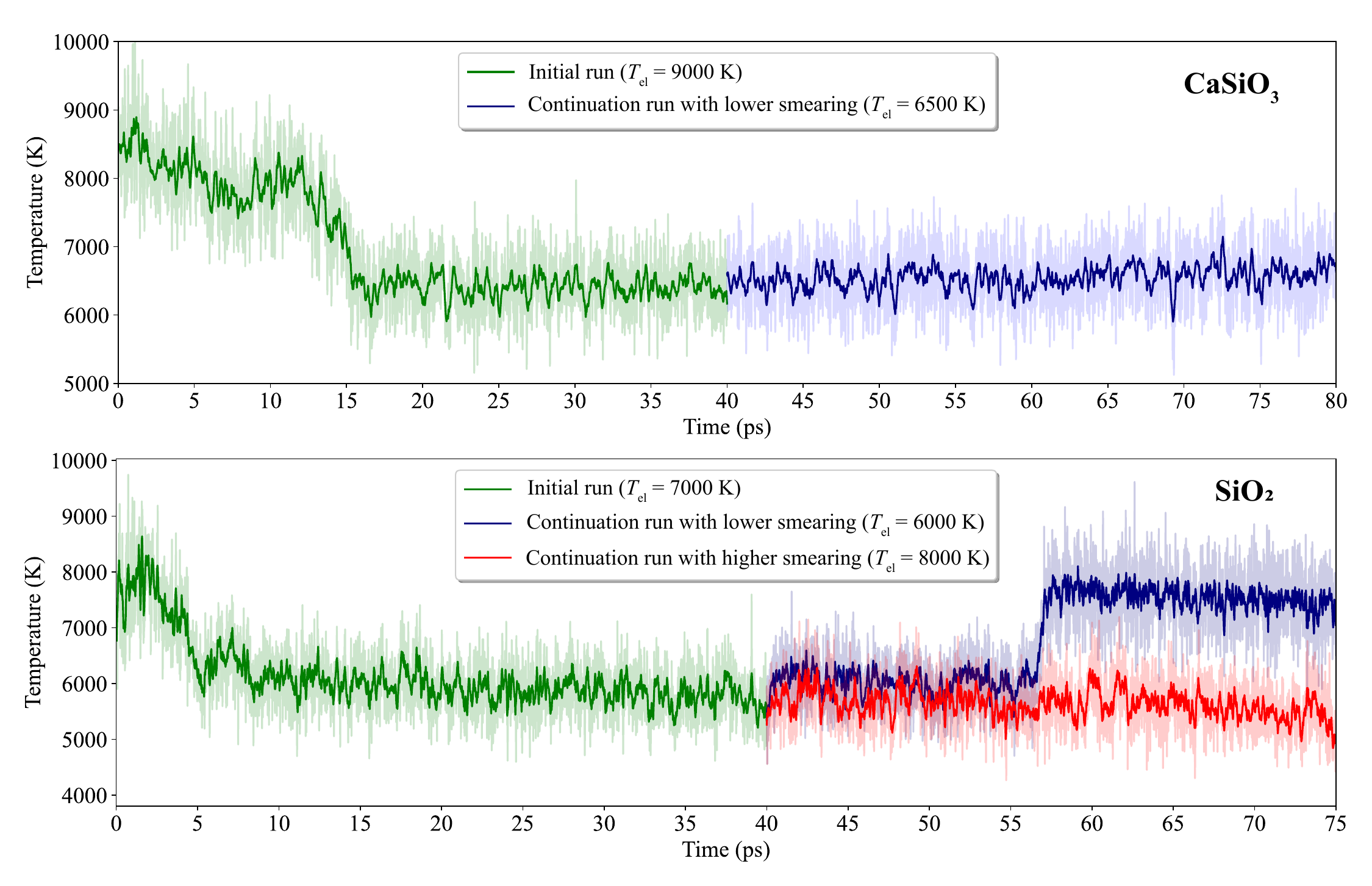}
    \caption{Blue and red lines are continuation runs after about 40 ps of simulations (green lines). These continuation runs restarted with the same atomic positions and forces as the last step of the initial (green) run, but with different electronic temperatures. For CaSiO$_3$, the MD run was initially launched with $T_{\text{el}}$ = 9000 K (green line) and then restarted with $T_{el}$ = 6500 K (blue line). For SiO$_2$, we started with $T_{\text{el}}$ = 7000 K (green line) and then continued with $T_{\text{el}}$ = 6000 K (blue line) and 8000 K (red line).}\label{propEl}
\end{figure*}

To further understand the role of electronic entropy on the Hellmann-Feynman dynamics we can follow a similar strategy as in Ref.~\cite{Hernandez2022} by adjusting the electronic entropy along the MD trajectory to match the new average ionic temperature after transition to the liquid state. 
We thus pick one of the  MD runs with $T_{\text{el}} \approx T_{\text{h}}$  and a total energy which is marginally higher than $E_{\text{h}}$. Using CaSiO$_3$ as an example, we expect that the average liquid temperature is close to the equilibrium melting temperature estimated from Eq. 3. 
Indeed,  $\langle T \rangle$ after melting is 6408 K at $\sim$ 105 GPa, which is within the error-bars of $T_{\text{m}}$ (6410 K $\pm$ 21 K). The simulation was then restarted after 25 ps of propagation in the liquid state with a new electronic temperature chosen to match the average liquid temperature. In Fig.~\ref{propEl} (top panel) we show the temperature evolution of this restarted run (blue).  The total energy increased by about 1\% accompanied by a temperature increase of about 100 K to 6517 K (labelled as "adjust" in Table 1). The $\langle T \rangle _{\text{liq}}$ from this relaunched MD run (blue line) is higher and lower respectively than the melting temperature reported in Table \ref{table} for $T_{\text{el}} = 9500$ K and $T_{\text{el}} = 6500$ K respectively. This  confirms that   $T_{\text{el}} \approx T_{\text{h}}$ and $T_{\text{el}} \approx T_{\text{m}}$ represent reasonable upper and lower bounds to the "true" melting temperature



Fig.~\ref{propEl} for SiO$_2$ also demonstrates that changes in electronic entropy can have a substantial impact on the ionic dynamics. Here we relaunch a simulation by changing the electronic temperatures from 7000 K (green line) to $T_{\text{el}}$ = 6000 K (blue line) or $T_{\text{el}}$ = 8000 K (red line). Whereas the MD run with $T_{\text{el}}$ = 6000 K {\it recrystallized} after about 15 ps of propagation in the liquid state the simulation with  $T_{\text{el}}$ = 8000 K remained stable in the liquid phase until the run was terminated after more than 35 ps.

\section{Conclusions}

In this work, we used the Z method together with {\it ab initio} Born-Oppenheimer molecular dynamics to calculate the melting temperature for SiO$_2$ and CaSiO$_3$ at outer core and lower mantle conditions respectively with simulation boxes containing $\sim$ 100 atoms only. The calculated melting temperature for CaSiO$_3$ is in excellent agreement with results from previously reported two-phase co-existence calculations and thermodynamic integration and is substantially higher than that calculated in a previous {\it ab initio} study using the Z method~\cite{Braithwaite2019}. A possible explanation for this discrepancy has been discussed. The calculated melting temperature for SiO$_2$ is also in overall good agreement with previous computational work carried out at similar pressure and temperature suggesting that the estimated equilibrium melting curve reported in Ref.~\cite{Millot2015} may be too flat. 

One of the great advantages with the Z method compared to other popular methods to melting, such as two-phase simulations and thermodynamic integration, is that the equilibrium melting temperature can be accurately calculated using small or modest-sized simulation boxes.  This is because melting near $T_{\text{h}}$ is in general triggered by the formation of defects and small liquid clusters which can be embedded in a small simulation box~\cite{Davis2019}. 
Moreover, since the melting temperature is obtained from the relationship between $T_{\text{h}}$ and $T_{\text{m}}$ using Eq.\ref{eq1}, the Z method avoids the calculation of free energies which can be extremely tedious and expensive. 
The Z method can easily be implemented and interfaced with popular {\it ab initio} simulation software such as VASP~\cite{VASP93, VASP94} and is embarrassingly parallelizable.

In spite of many appealing advantages compared to other methods to melting, the Z method appears to be hampered by some artificial features which needs to be much better addressed.  
One of these, investigated here, is the choice of electronic entropy in Born-Oppenheimer MD simulation carried out in the microcanonical ensemble when  $ K + \Omega $  is a conserved and hence $T_{\text{el}}$  is kept fixed along the Hellmann-Feynman trajectory.  Since melting is accompanied by a large drop in temperature we need to quantify the errors introduced due to the choice of electronic temperature/entropy. We, therefore, compare the melting temperature calculated using  $T_{\text{el}} \approx T_{\text{h}}$ and $T_{\text{el}} \approx T_{\text{m}}$ since these represent reasonable upper and lower bounds to the "true" melting temperature.  For SiO$_2$ and CaSiO$_3$ at high pressure and temperature, the difference in the calculated $T_{\text{m}}$ with $T_{\text{el}} \approx T _{\text{h}}$ and $T_{\text{el}} \approx T _{\text{m}}$ is about 200-300 K. Although these discrepancies are only a few percent of the melting temperature, they are not negligible for the accurate calculation of melting temperature and are important to take into account for a critical assessment of the Z method implemented together with Born-Oppenheimer MD in the $NVE$ ensemble with a constant electronic temperature.

\begin{acknowledgments}
We thank Prof. Dario Alf\`e for the thoughtful discussion. The Centre for Earth Evolution and Dynamics is funded by the Research Council of Norway through its Centre of Excellence program (Grant 223272). We acknowledge financial support from the Research Council of Norway through its Centres of Excellence scheme, project number 332523 (PHAB). Computational resources were provided by the Norwegian infrastructure for high-performance computing (NOTUR, Sigma-2, Grants NN9329K and NN2916K). 
\end{acknowledgments}

\nocite{*}

\bibliography{apssamp}

\clearpage
\beginsupplement

\title{Influence of electronic entropy on Hellmann-Feynman forces in {\it ab initio} molecular dynamics with large temperature changes 
\\ Supplementary Information}
\author{Ming Geng $^{1}$}
\author{Chris E. Mohn$^{1,2,}$}
\email{chrism@kjemi.uio.no}
\affiliation{
$^{1}$The Centre for Earth Evolution and Dynamics (CEED)  and Centre for Planetary Habitability (PHAB), University of Oslo, N-0315 Oslo, Norway
}
\affiliation{
$^{2}$Department of Chemistry and Center for Materials Science and Nanotechnology, University of Oslo, Oslo 0371, Norway
}
\maketitle

\begin{figure*}
    \includegraphics[width=0.9\textwidth]{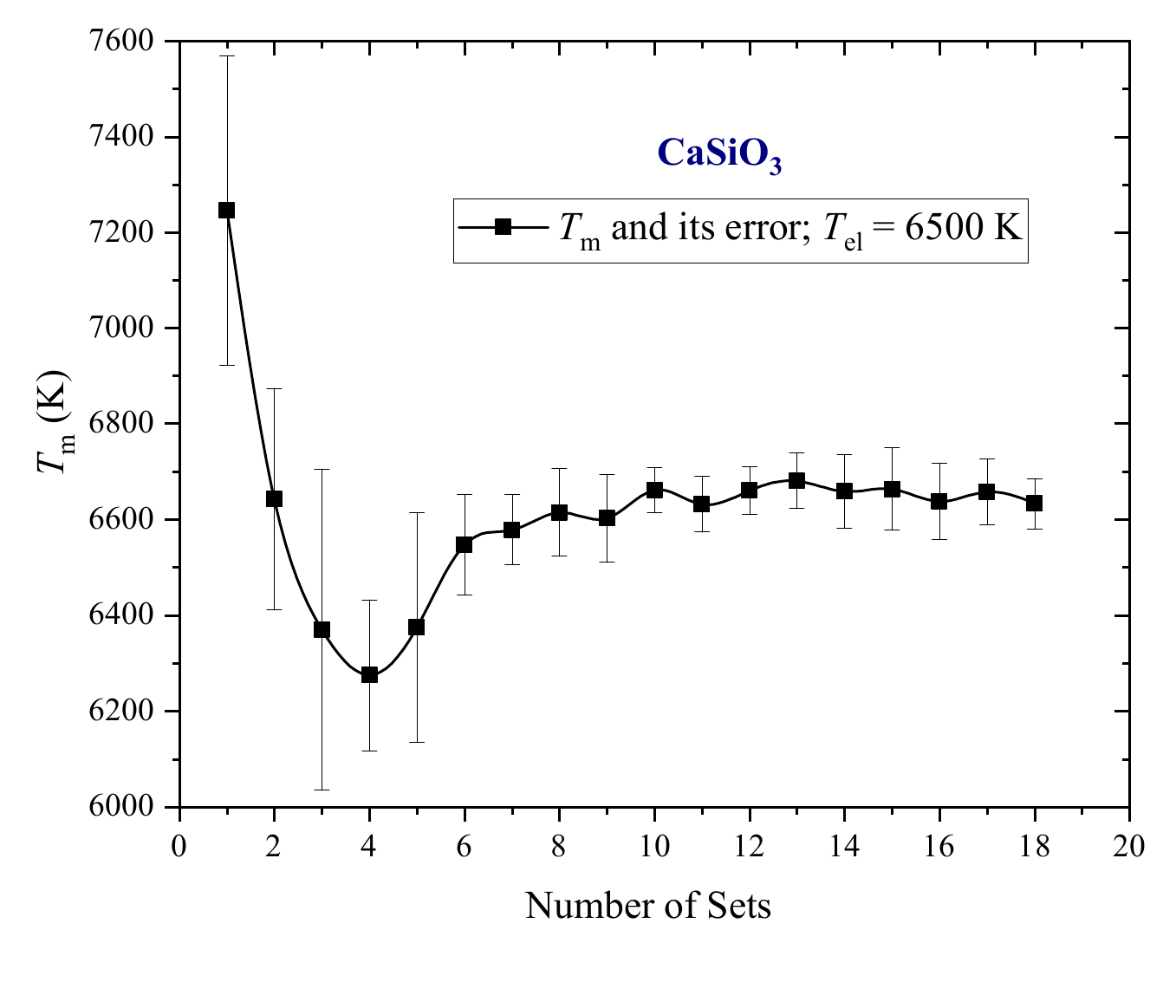}
    \caption{Convergence of T$_{\text{m}}$ with number of MD runs for CaSiO$_3$. Each "set" of MD runs contains three different initial temperatures.}\label{ConvergenceCaSiO3}
\end{figure*}

\begin{figure*}
    \includegraphics[width=0.9\textwidth]{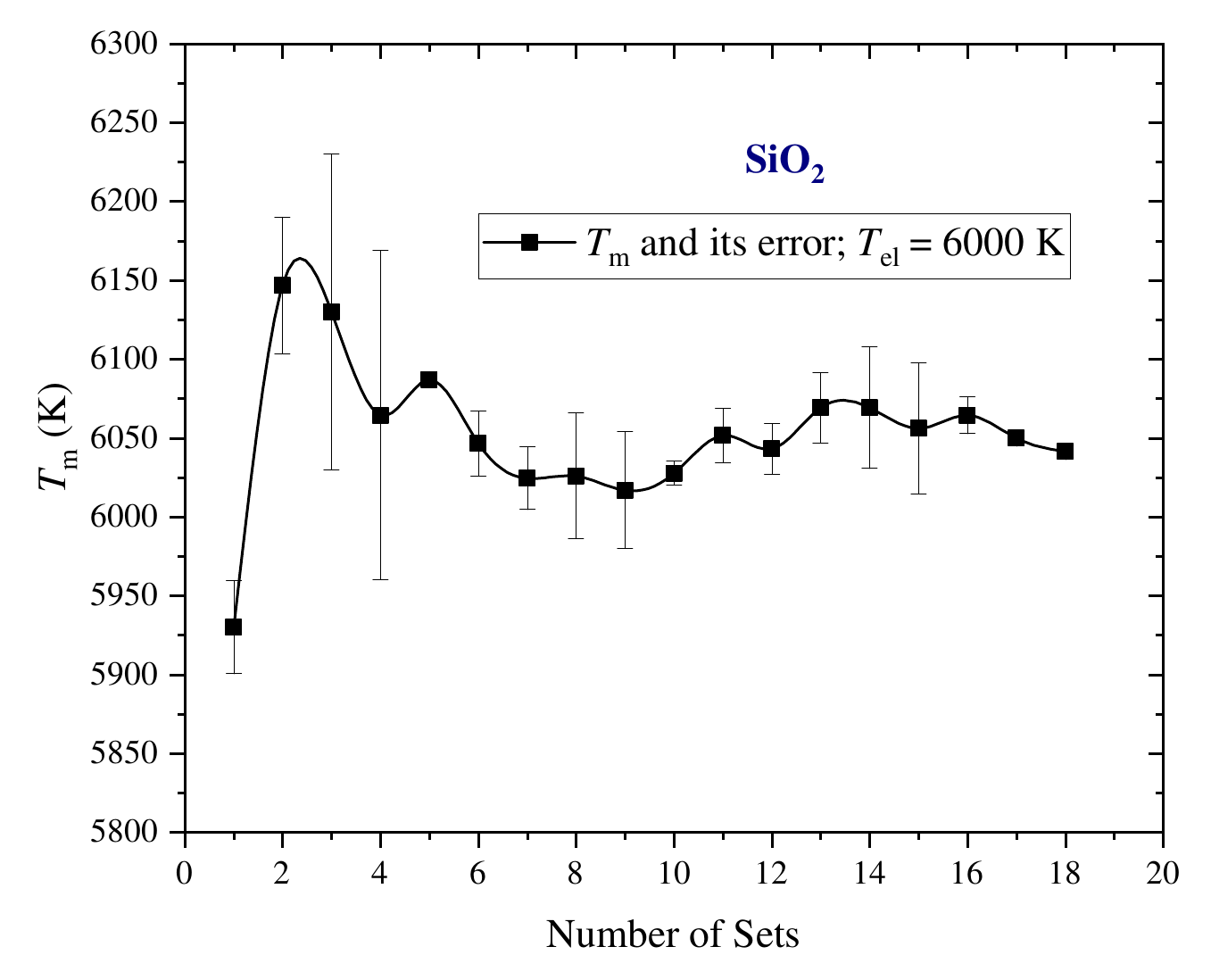}
    
    \caption{Convergence of T$_{\text{m}}$ with number of MD runs for SiO$_2$. Each "set" of MD runs contains three different initial temperatures.}\label{ConvergenceSiO2}
\end{figure*}

\subsection{Distribution of waiting times and mean errors in the calculated melting temperature}
The use of a waiting time analysis via Eq.~\ref{Eqwait} provides a powerful implementation of the Z method, but many MD runs is still needed for the precise calculation of $T_{\text{m}}$. In Figs. S1 and S2 we show the convergence of the melting temperature with number of sets where each "set" contains one MD run from each initial temperature. This shows that about 10 simulations of each set (about 30 MD runs in total) gives mean errors of less than 100 K. However many of these runs lasted for more than 200 ps and more MD runs would have been beneficial, in particular close to the melting temperature. Although the stochastic nature of melt nucleation and long tails in the waiting time distribution remains a challenge to the Z method~\cite{Alfe2011, Davis2019}, the emergence of machine learning could provide a powerful future approach to improve statistics~\cite{VASP93, VASP94}.

The stochastic nature of melting is illustrated as an inset in Fig. S3. Here we show the time evolution of a number of MD runs with the same total energy, $E$, which is slightly above $E_{\text{h}}$. As seen, some calculations melted fairly quickly ($<$ 10 ps)  whereas others remained stable for more the 200 ps. Although we do not have sufficient statistics to capture the functional form of the distribution of $\tau $, analysis carried out by Davis {\it et al}~\cite{Davis2019}  showed  - using a classical Lennard-Jones potential to model melting of Ar - that the distribution of waiting times appears to deviate markedly from that of an exponential decay. 

\begin{figure*}
    \includegraphics[width=0.9\textwidth]{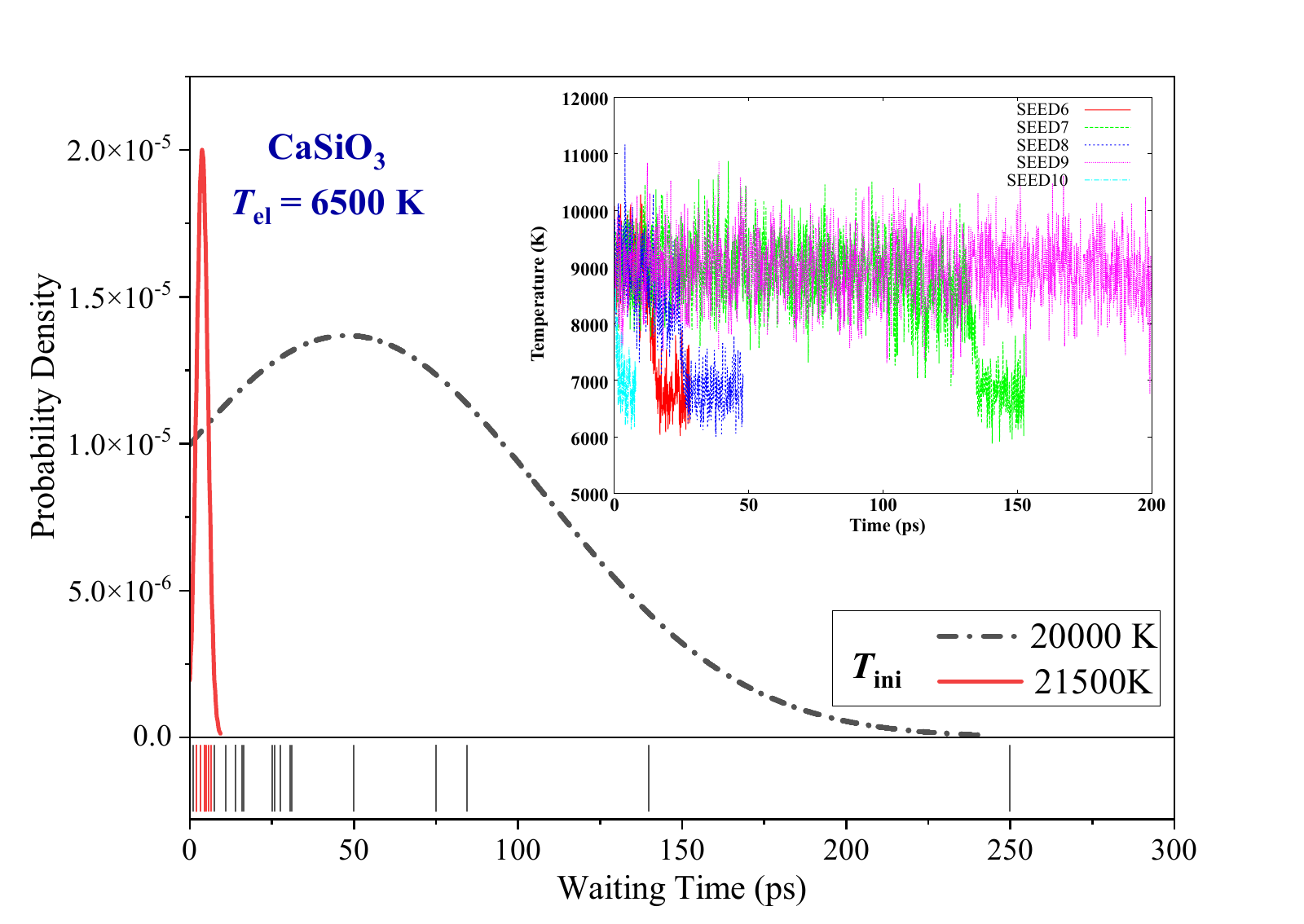}
    \caption{Distribution of waiting times for melting and examples of $NVE$ MD runs of CaSiO$_3$  with $T_{\text{ini}}$ = 20000 K and $T_{\text{el}}$ = 6500 K.  A normal-distribution is fitted to the distribution probability density. The inset figure shows five different seeds of the temperature evolution in an NVE MD runs with $E > E_{\text{h}}$ for CaSiO$_3$. In all runs the atoms were located at their equilibrium positions and the velocities were drawn from a Maxwell-Boltzmann distribution.}\label{disttauCaSiO3}
\end{figure*}

\begin{figure*}
    \includegraphics[width=0.95\textwidth]{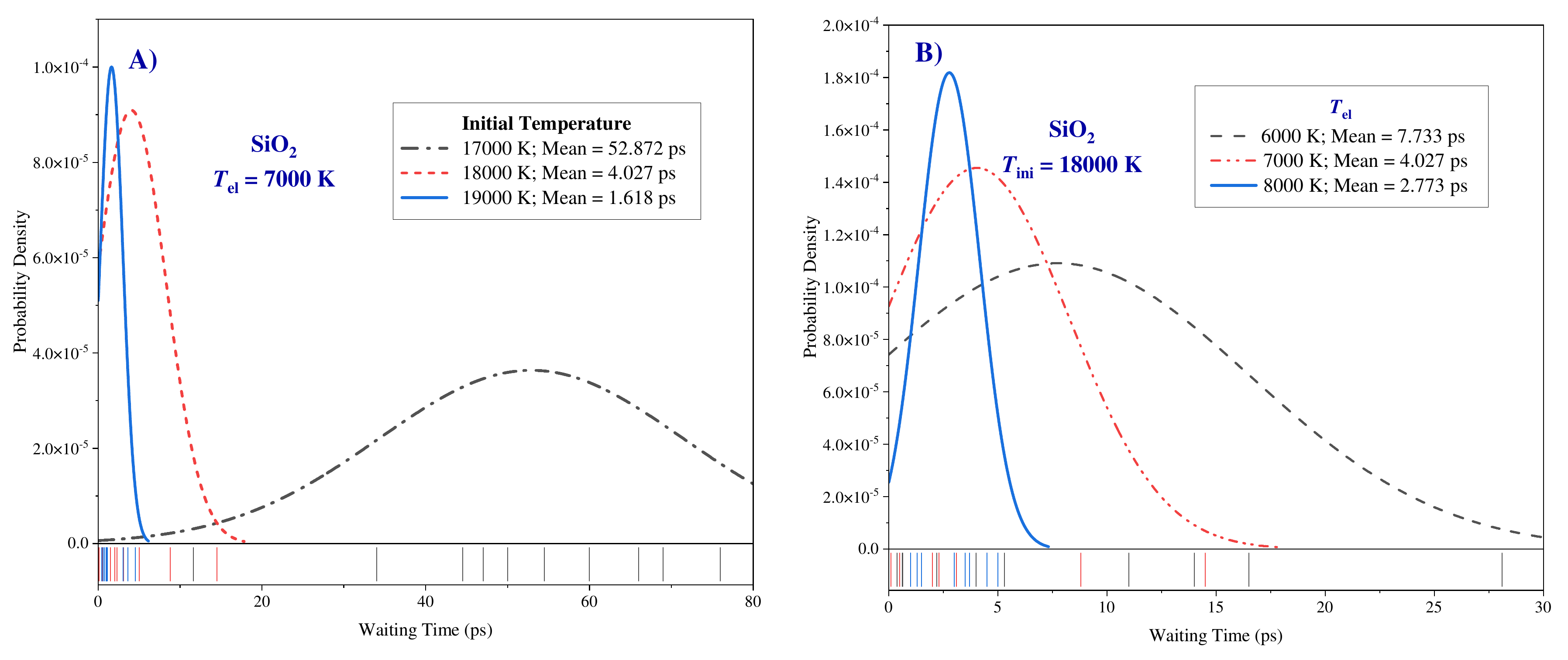}
    \caption{Distribution of waiting times for melting of SiO$_2$.  A  function is fitted to the distribution probability density. In the left figure we show distributions from runs with the same electronic temperature but different initial temperature and in the right one the initial temperatures are the same but the  electronic temperatures are different. In all runs, the atoms were located at their equilibrium positions and the velocities were drawn from a Maxwell-Boltzmann distribution. }\label{disttauSiO2}
    \label{fig:SiO2_waiting_distribution}
\end{figure*}

Although exponential decays describe many dynamical atomistic processes such as hopping events in solids~\cite{Mohn1CuI, MohnBi2O3}, the time taken from random fluctuations that trigger melting to when melting is observed will always take a certain time~\cite{Davis2019}. This suggests that $\tau$ deviates from an exponential decay at short times~\cite{Davis2019}. 
In Fig.~\ref{disttauCaSiO3} and ~\ref{disttauSiO2}, we plot the distribution of the waiting times and possible (fitted) distributions
for CaSiO$_3$ and SiO$_2$ respectively. The plots confirm that the melting time is sensitive to initial temperature and electronic entropy as discussed above. The distributions of $ \tau  $ with $T_{\text{el}}$  close to $T_{\text{m}}$ are wide and with standard deviations which are on the size of the average waiting time. For CaSiO$_3$ runs (with $T_{\text{ini}}$ = 20000 K and  $T_{\text{el}}$ = 6500 K), for example, the mean waiting time is $\sim$ 50 ps with standard deviations of similar size when taken from a normal distribution as shown in Fig.~\ref{disttauCaSiO3}.

If we assume that the melting temperature can be precisely determined by Eq.~\ref{Eqwait}, a fruitful strategy to minimize the computational cost in a waiting time analysis would be to choose the total energies/initial temperatures such that $\langle \tau \rangle < 10$ ps and $\tau > \tau_{\text{melting}} \approx$ 1 ps. The upper bound ensures that we minimize lengthy MD runs, whereas the lower bound, $\tau _{\text{melting}}$, ensures that the waiting time is markedly longer than the time it takes from liquid
precursors is formed until the entire system has melted which, in our cases where rather small simulations boxes have been used, are typical $ \sim $ 1 ps. (see Fig.~\ref{fig1} in the main text). If the waiting time is shorter than the melting time, it is difficult to precisely identify the waiting time in a given run and hence the error bars in $\langle \tau \rangle ^{-1/2}$ are large (as seen in Fig. 4 in the main text at short waiting times).


\subsection{Solid-liquid oscillations}

\begin{figure*}
    \includegraphics[width=0.9\textwidth]{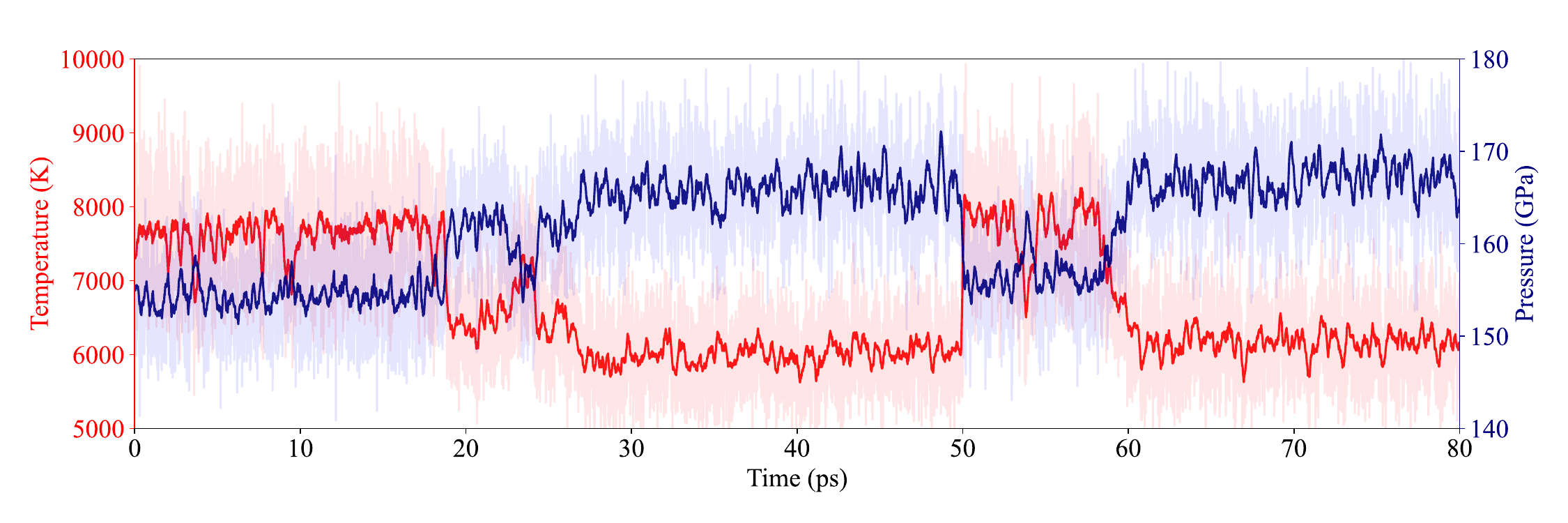}
    \caption{Temperature and pressure evolution in an SiO$_2$ MD run  where T$_{\text{ini}}$ = 17000 K and $T_{\text{el}}$ = 8000 K. Transition between liquid and solids take place at 25 ps, 50 ps and 60 ps. Note the overlapping fluctuations in temperature/pressure along solid and liquid states.}\label{fig6}
\end{figure*}

Although solid-liquid transitions may be triggered by small changes in electronic entropy, refreezing can also occur spontaneously in $NVE$ runs when the temperature fluctuations are large. That is, when there is an overlap between liquid and solid temperature distributions as discussed in Ref.~\cite{Alfe2011}.
Fig.~\ref{fig6} illustrates solid-liquid oscillations for SiO$_2$. Here the temperature drops after about 20 ps as the system melts, followed by a 30 ps evolution in the liquid state before the system rapidly recrystallizes. After 10 ps in the solid state, it melts again and remains stable in the melt for more than 20 ps before the run is terminated.
Since, however, the temperature fluctuations scale as $1/\sqrt{N}$ (where $N$ is the number of atoms in the simulation box), the overlap between the solid and liquid temperatures decreases with increasing box size and hence the frequency of the solid-liquid alternation decreases. Oscillation on a time scale of nanoseconds or less is therefore usually only seen in MD runs carried out in small simulation boxes ($\sim$ 100 atoms). 

An interesting implication of such solid-liquid oscillation is that it is possible to estimate the melting temperature by targeting the total energy in which the system spends an equal amount of time in the solid and liquid phases. In this case, the entropy of the two phases are equal and hence the melting temperature can be extracted directly from the average temperature in the liquid state~\cite{Alfe2011}.  
However, in our case of SiO$_2$ melting at $\sim$ 160 GPa, extremely long runs - probably on the size of tens of nanoseconds -  would be needed to collect sufficient statistics. In addition, analysis of all trajectories shows that liquid-solid transition is rare in $NVE$ MD runs in SiO$_2$, and for CaSiO$_3$ we did not observe any re-crystallization events even though the size of the simulation box size is modest (135 atoms). The emergence of machine learning with "on the fly" construction of force-fields along the MD trajectory could provide a powerful future implementation of the Z method to improve statistics. This would enable one to explore such oscillations on much longer time-scales than that possible with conventional BOMD simulations.

\end{document}